\documentclass[vecphys]{svmult}

\usepackage{makeidx}         
\usepackage{graphicx}        
\usepackage{multicol}        
\usepackage[bottom]{footmisc}

\makeindex             

\def\Deg{${}^\circ$\llap{}}

\begin{document}

\title*{Studying galaxy formation and evolution from Local Group galaxies.}
\author{Carme Gallart\inst{1}}
\institute{Instituto de Astrof\'\i sica de Canarias.
\texttt{carme@iac.es}}
%
%
\maketitle
\index{Gallart, C.}
\begin{abstract}
In this contribution I present the main research activities of the IAC
 ``Stellar Populations in Galaxies'' research Group, with emphasis on the
subtopics directly related with the study of the evolution of nearby
galaxies. In particular, I discuss preliminary results of ongoing
research on the Magellanic Clouds using deep ground-based
observations, and on a sample of isolated Local Group dwarf galaxies,
using data from the ACS on board the HST. Future plans with the GTC
are discussed.
\end{abstract}

\section{Introduction}
\label{intro}

\noindent The process of galaxy formation and evolution is driven by
two sets of parallel mechanisms: stellar formation, evolution and
death, which drive the evolution of the stellar populations and of
the gas and metal content of the galaxy, and the mass assembly
process, which determines its morphological type and dynamical
evolution, and which in turn may induce star formation. In the nearest
objects, and in particular in those that can be resolved into stars
like Local Group galaxies, these mechanisms can be studied in great
detail. We are conducting a comprehensive study of Local Group
galaxies using a number of complementary tools to shed light on
these two main mechanisms that determine galaxy formation and
evolution, namely:

\noindent {\it 1. The star formation history (SFH), and its influence on 
galaxy evolution, through:} i) The study of deep colour--magnitude
diagrams (CMDs) of each galaxy; ii) the spectroscopic abundances of
resolved stars; iii) the analysis of the properties of their variable
stars; iv) the study of the Milky Way Cluster system.

\noindent {\it 2. The mass assembly and the dynamical evolution of each 
system, through:} i) The stellar population gradients and kinematics
of stars of different ages; ii) the dynamics of the Local Group and
the influence of interactions on galaxy evolution.

This project is designed  to make the best use of the GTC
and its first-light instruments, especially OSIRIS. In this
contribution, I will present examples of ongoing research in several
of the aspects quoted above.

\section{The star formation history from deep colour--magnitude diagrams.}
\label{cmd}

The CMD, and in particular those reaching the oldest main-sequence
turnoffs, is the best tool for retrieving in detail the SFH of a
stellar system.  In this case, information on the distribution of ages
and metallicities of the stars present in the galaxy can be obtained
directly from stars on the main sequence, which is the best understood
phase of stellar evolution from the theoretical point of view. It is
also the one in which stars are more separated in colour and magnitude
as a function of age. The range of ages and metallicities present can
be determined through comparison with theoretical isochrones. To
quantitatively determine the SFH, it is necessary to compare the
observed density distribution of stars with that predicted by stellar
evolution models (see \cite{gal05}).

Our group has been traditionally dedicated to the derivation of SFHs
from deep CMDs, through comparison of observed and synthetic
CMDs (e.g.\ \cite{gal99}, \cite{apa01}, \cite{car02}, \cite{apa04}; see
also http://iac-star.iac.es/iac-star/). We are currently involved in
two major programmes in this regard. The first aims at providing
spatially resolved SFHs for both Magellanic Clouds, using ground-based
observations. These, however, have produced CMDs of quality comparable
(but covering a much larger area) to CMDs obtained in the central
regions of both objects using the WFPC2 on HST (e.g.\ \cite{hol99},
\cite{sme02}). The second is devoted to obtaining detailed SFHs for a
sample of isolated Local Group dwarf galaxies, using ACS CMDs reaching
the oldest main-sequence turnoffs.

\begin{figure}
\centering
\includegraphics[height=12.5cm]{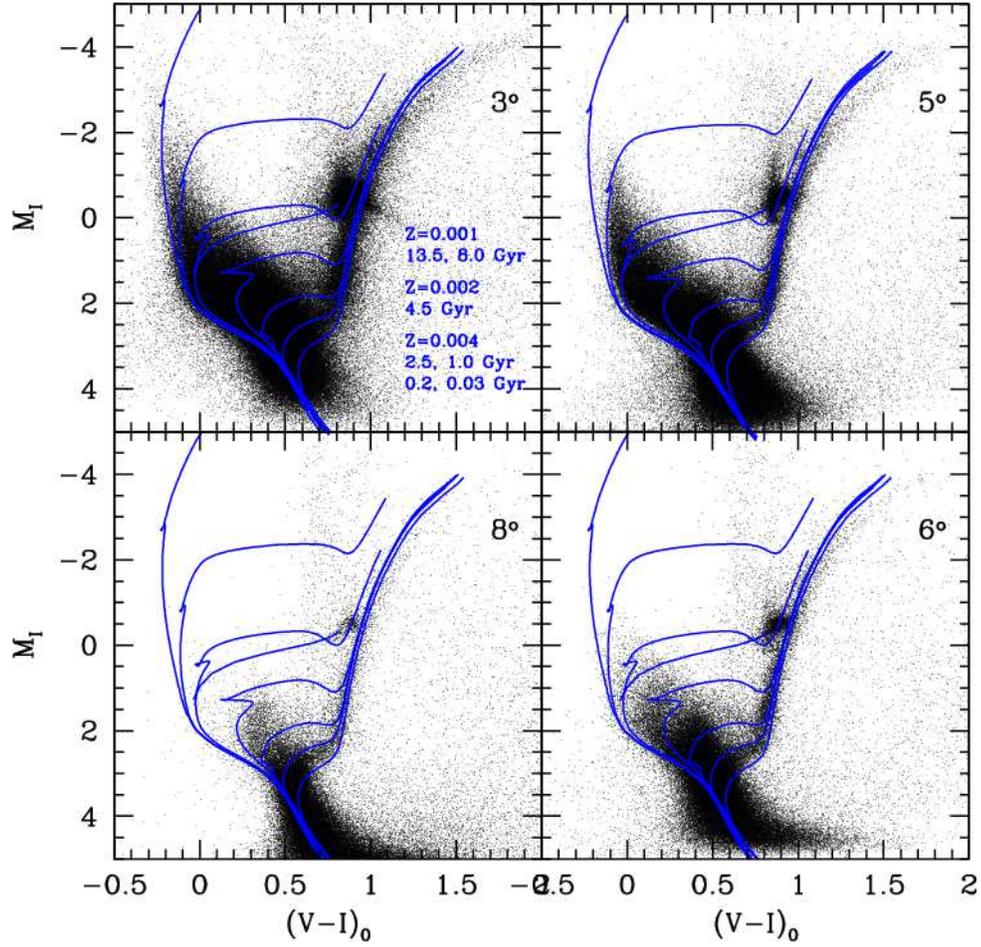}
%
%
\caption{CMDs for the four LMC fields observed with the MOSAIC Camera at 
the 4 m CTIO, located at $\simeq$ 3\Deg, 5\Deg, 6\Deg\ and 8\Deg\ from
the LMC centre (clockwise from top left, respectively). Isochrones
from \cite{pie04} (scaled solar, overshooting set), with the ages and
metallicities indicated in the labels, have been superimposed. The
locus of the zero-age horizontal-branch is that of the lowest
metallicity considered. A distance modulus of $(m-M)_0=18.5$ and
reddenings of $E(B-V)=0.1$, 0.05, 0.04 and 0.03 have been assumed to
transform the data to absolute magnitudes and colours. Determinations
of the SFH of each field are under way through comparison with
synthetic CMDs. }
\label{figlmc}       
\end{figure}

\subsection{The LMC} \label{lmc}

In the case of the LMC (and as part of the PhD thesis of
I. Meschin), we have observed 12 half degree fields, six to the
north and three each to the east and west of the galaxy centre,
using the MOSAIC camera at the 4 m CTIO and the WFI at the 2.2 m in La
Silla. These fields sample galactocentric distances (from 3\Deg\ to
10\Deg, or 2.6 to 8.8 kpc) not explored before to this photometric
depth, i.e.\ with CMDs reaching the oldest main-sequence
turnoffs. Through comparison with synthetic CMDs, these data will
allow us to obtain detailed SFHs for all these fields and characterize
the population gradients present in the galaxy.  Figure~\ref{figlmc}
shows the CMDs for four fields observed to the north of the
galaxy, obtained from the MOSAIC data. Isochrones from \cite{pie04} in
a suitable range of age and metallicity have been superimposed. It can
be noticed that the age of the youngest population varies from field
to field, in the sense that star formation has proceeded down to more
recent epochs toward the galaxy centre: while in the fields situated
at 3\Deg\ and 5\Deg, star formation has basically continued to the
present time, the field at 6\Deg\ seems not to have stars younger than
$\simeq 200$ Myr. Finally, the field at 8\Deg\ formed the bulk of its
stars before 2.5 Gyr, with some residual star formation up to 1.5 Gyr
ago. The presence of an important intermediate-age population in this
field, together with the fact that the surface brightness profile of
the LMC remains exponential to this large galactocentric radius and
shows no sign of disk truncation, led \cite{gal04b} to conclude that
the LMC disc extends (and dominates over a possible stellar halo) out
to a radius of at least 7 kpc.

\subsection{The SMC} \label{smc}

For the SMC (and as part of the PhD thesis of N. No\"el), we have
similar CMDs for 13 smaller fields observed with the
100$^{\prime\prime}$ telescope at LCO (\cite{noe06}). These fields are
distributed in different parts of the SMC such as the ``Wing'' area
and to the west and south, and in a range of galactocentric distances
(from $\simeq$ 1\Deg\ to 4\Deg, or 1 to 4.1 kpc). Several studies
(e.g.\ \cite{zar00},
\cite{cio00}) have found that the SMC intermediate-age and old
population has a spheroidal distribution, and that the asymmetric
appearance of the SMC is primarily caused by the distribution of young
stars. With our deeper data, we can shed new light on the age
distribution of these structures. In particular, we confirmed that the
underlying spheroidal population is composed of both intermediate-age
and old stars, and found that its age composition does not show strong
galactocentric gradients. The three fields situated in the ``Wing''
region show very active current star formation, but only the one
closer to the centre seems to present a substantial enhancement in
recent star formation with respect to a constant SFR($t$). The fields
corresponding to the western side of the SMC present a much less
populated young main sequence as compared with those on the east side,
even at similar galactocentric radius, with signs of a greatly
diminished SFR($t$) from 2 Gyr ago to the present time. As in our LMC
study, none of the studied fields, out to a galactocentric radius of
4\Deg (or 4.2 kpc), is dominated by an old stellar population.

\subsection{Isolated Local Group dwarf galaxies}\label{lcid}

We are participating in two HST-ACS programs (P. ID: 10505, P.I. Gallart;
P.ID: 10590, P.I. Cole) with a total of 113 awarded orbits, in
addition to an HST-WFPC2 program (P.ID: 8706, P.I. A. Aparicio), to
obtain CMDs reaching the oldest main-sequence turnoffs in 6 isolated
Local Group galaxies (two dIrr galaxies: Leo~A and IC1613, the two
isolated dSph galaxies discovered so far in the Local Group: Cetus and
Tucana, and two transition type dIrr/dSph galaxies: LGS3 and Phoenix).

\begin{figure}
\centering
\includegraphics[height=12.5cm]{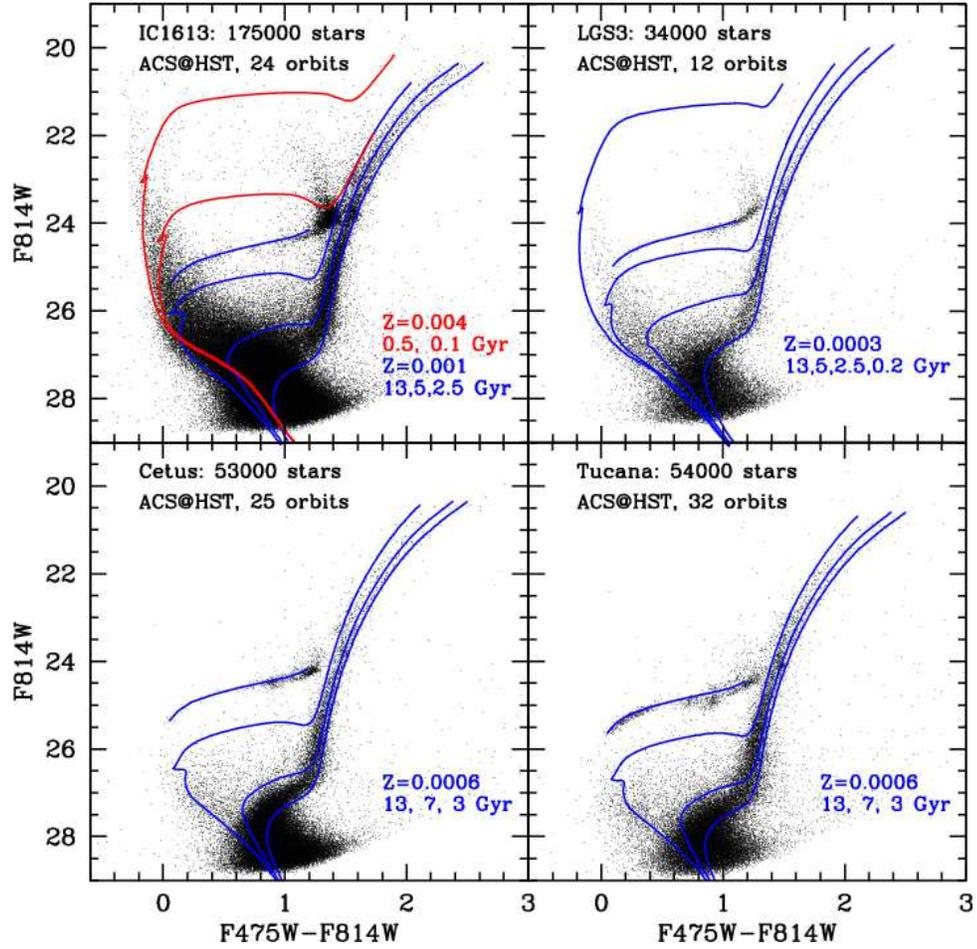}
%
%
\caption{CMDs obtained with the ACS on board HST for four isolated
Local Group dwarf galaxies. In order to give a first indication of the
range of ages and metallicities present in each galaxy, isochrones
from \cite{pie04} (scaled solar, overshooting set), with the ages and
metallicities indicated in the labels, have been superimposed. The
locus of the zero-age horizontal-branch has also been
represented. Distance moduli of $(m-M)_0=$ 24.4, 24.0, 24.45 and 24.7
and reddenings of E(B-V)=0.04, 0.05, 0.03 and 0.03 for IC1613, LGS3,
Cetus and Tucana, respectively, have been adopted to transform the
isochrones to the observational plane.  Determinations of the SFH of
each system are under way through comparison with synthetic CMDs.}
\label{figacs}       
\end{figure}

Figure \ref{figacs} shows the CMDs of four of the galaxies in the
sample. Note the variety of SFHs, as hinted at by the comparison with
selected isochrones from \cite{pie04}. This is the first
time that data of this high quality has been obtained for dwarf galaxies
beyond the Milky Way satellite system.  These data will allow us to
obtain detailed and accurate SFHs for all these systems, through
comparison with synthetic CMDs, and using additional constraints from
the characteristics of their variable star population (see Section
\ref{var} below, and the contribution by E. Bernard {\it et al.}  in these
Proceedings). 

The details of the early SFHs of tiny dwarf galaxies can shed light,
in particular, on the role in galaxy formation of the reionization
which occurred at high redshift. Isolated dwarfs are ideal probes
since their evolution is not complicated by environmental effects
owing to the vicinity of the Milky Way or M31.

\subsection{The study of stellar population gradients} \label{gradients}

In all these programmes, in addition to the derivation of accurate and
detailed SFHs, we pay special attention to the study of the stellar
population gradients. Their presence in dwarf galaxies, with the
youngest population concentrated towards their centre, is well known
(e.g.\ \cite{apa97}, \cite{apa00}, \cite{har01}). In the Milky Way
satellites, for which we have CMDs reaching the oldest main-sequence
turnoffs, the nature of these gradients can be investigated in detail;
here the difficulty is due to the large areas that need to be
surveyed. For example, in the case of the Magellanic Clouds, their
total extension and the presence or not of an old halo is still a
matter of debate. In the case of the more distant dIrr galaxies, the
actual nature of the outer older structure, and its extension, is
still uncertain due to the faintness of the stars that need to be
measured. Recent studies show the existence of both young and old
populations in the central parts of dIrr galaxies, while young stars
gradually disappear towards the outer regions. But they don't offer
enough proof of the actual nature of these extended structures, and in
particular, whether they represent a true halo population (i.e.,\ old
and tracing the initial conditions of galaxy formation). Recent
results for the LMC (\cite{gal04b}) and Phoenix (\cite{hid06}), using
CMDs that reach the oldest main-sequence turnoffs indicate that these
extended regions are not exclusively old, and that a smooth age
gradient exists from the centre of the galaxy to its outer parts. This
suggests an {\it outside--in} formation scenario, contrary to what
seems to happen to the discs of spiral galaxies.

With the CMDs of isolated dwarfs shown in the previous Section, it
will be possible to investigate in detail the age distribution of the
stellar populations in each galaxy as a function of radius, thus
sheding new light on the possible formation mechanisms. An
additional key diagnostic is the kinematics of stars
(e.g.,\ \cite{tol04}) of different ages, which will provide information
of the dynamical evolution of the galaxy, possibly indicating the
presence or otherwise of differentiated disc--halo structures, or the
presence of distinct kinematic entities, possibly originating in the
accretion of smaller systems (according to the predictions of
hierarchical galaxy formation models). This is a field still to be
explored for dwarf galaxies outside of the Milky Way satellite system,
and Flames at the VLT and OSIRIS at the GTC will be key instruments
for this purpose.

\section{Metallicities using the Ca II triplet.} \label{caii}

We have obtained (PhD, R. Carrera) a new calibration of the CaII
triplet strength in red giant branch (RGB) stars as a function of
metallicity, which is valid for a higher range of ages (13 $\le$
Age(Gyr) $\le$ 0.25) and metallicities ($-2.2 \le$ [Fe/H] $\le$ +0.5)
than previously published calibrations (see \cite{col04} for the most
recent one).  This calibration has been used to obtain metallicities
for a large number of stars in different fields of the LMC and the SMC
(see the contribution by R. Carrera {\it et al.} in these Proceedings
for details).

With the GTC and OSIRIS we plan to extend this type of work to the
remaining galaxies in the Local Group, situated at a distance of
$\simeq$ 1 Mpc. Candidate RGB stars are in the magnitude range
$I\simeq$21--21.5.  They will densely populate the OSIRIS field of
view ($\simeq$30 stars per sq.\ arcminute) to allow us efficient use of
multiobject spectroscopy. The high brightness of the sky in the CaII
triplet region implies that the possibility of micro-slit
nod-and-shuffle will be important for this project.

\section{Variable stars} \label{var}

Variable stars can complement the information offered by CMDs to
interpret the stellar populations of a galaxy. In particular, RR Lyrae
reveal the presence of a very old ($\simeq$10 Gyr) stellar population,
while short-period classical Cepheids and anomalous Cepheids are
tracers of populations up to a few hundred Myr old and a few Gyr old
respectively (\cite{gal04a}). Using variable stars as stellar
population indicators is especially important when it is not possible
to obtain CMDs reaching the oldest main-sequence turnoffs.

The ACS data mentioned in Section~\ref{cmd} is also excellent for
obtaining a census of the variable star population in each galaxy of the
sample. Such a study is already under way (see the contribution by
E. Bernard {\it et al.}\ in these Proceedings). The good sensitivity and
relatively large field of OSIRIS at the GTC will allow us to carry on
a systematic characterization of the variable star populations in
Local Group galaxies (both isolated dwarfs and M31 dSph companions,
and some strategic fields in the large M31 and M33 spirals). Such
surveys will complement in a key way our ACS imaging project: with ACS,
only a handful of galaxies will be studied, and in some cases only
part of their total extent will be covered. To find RR Lyrae stars,
the most challenging and interesting part of this project, we need to
reach g$^{\prime}\simeq 25.5$ in relatively short exposure times. OSIRIS will
allow us to do that in 10--15 minutes. In addition, its field of view
is very well suited to cover most Local Group dwarf galaxies in one to
a few fields.

\acknowledgement Support for this project is provided by the IAC 
(Project 3I1902), the Spanish Ministry of Science and Technology
(AYA2004-06343), and the European Structural Funds. The {\it Stellar
populations in Galaxies} group at the IAC is currently composed of
A. Aparicio, E.J. Bernard, R. Carrera, I. Drozdovsky, A. Mar\'\i
n-Franch, I.P. Meschin, M. Monelli, N.E.D. No\"el, A. Rosenberg and
myself. I thank the co-investigators of the projects discussed in this
paper for allowing me to show results in advance of publication. In
particular (and in addition to the IAC Group members quoted above)
F. Pont, E. Hardy, P. Stetson and R. Zinn (LMC project), E. Costa and
R. M\'endez (SMC project), and the LCID Team (Local Constraints from
Isolated Dwarfs Team: A. Aparicio, E.J. Bernard, G. Bertelli,
S. Cassisi, A.A. Cole, P. Demarque, A. Dolphin, I. Drozdovsky,
H.C. Ferguson, L. Mayer, M.L. Mateo, M. Monelli, J. Navarro,
S.L. Hidalgo, F.J. Pont, E.D. Skillman, P.B. Stetson \& E. Tolstoy).

%
%
%
%
%
%
%

%
%
%
%
%
%
%


\printindex
\end{document}